\documentclass[aps,preprintnumbers,amsmath,amssymb,nofootinbib,preprint]{revtex4}
\usepackage[latin9]{inputenc}
\usepackage{hyperref}
\hypersetup{
    colorlinks,
    citecolor=blue,
    filecolor=blue,
    linkcolor=blue,
    urlcolor=blue
}

\date{\today}
\usepackage{physics}
\usepackage{amsmath,amsthm,amssymb}
\usepackage{xcolor}
\usepackage{natbib}
\usepackage{graphicx}
\usepackage{bm}

\begin{document}
\title{Classical Entanglement and Entropy}

\author{Haowu Duan}
\affiliation{North Carolina State University, Raleigh, NC 27695, USA}
\author{Alex Kovner}
\affiliation{Physics Department, University of Connecticut, 2152 Hillside Road, Storrs, CT 06269, USA}
\author{Vladimir V. Skokov}
\affiliation{North Carolina State University, Raleigh, NC 27695, USA}
\affiliation{RIKEN-BNL Research Center, Brookhaven National Laboratory, Upton, NY 11973, USA}

\begin{abstract}
Motivated by recent discussions of entanglement in the context of high energy scattering, we consider the relation between the entanglement entropy of a highly excited state of a quantum system and the classical entanglement entropy of the corresponding classical system. We show on the  example of two weakly coupled harmonic oscillators, that the two entropies are equal. Quantum mechanically, the reduced density matrix which yields this entropy is close to the maximally entangled state. We thus observe that the nature of entanglement in this type of state is purely classical.
\end{abstract}

\maketitle
\newpage

\section{Introduction}
Quantum entanglement is a fundamental concept in quantum information theory. Recently interest in quantum entanglement has also expanded in other areas of theoretical physics~\cite{Klco:2021lap}. Our interest the subject stems from the possible relevance of this phenomenon to high energy hadronic scattering, in particular collisions involving heavy nuclei and Deeply Inelastic Scattering (DIS)~ \cite{Robin:2020aeh,Beane:2019loz,Armesto:2019mna,Neill:2018uqw,Kovner:2018rbf,Hagiwara:2017uaz,Hatta:2015ggc,Dvali:2021ooc,Ehlers:2022oal,Kou:2022dkw,Ramos:2022gia,Dumitru:2022tud}. 

In the context of heavy ion collisions and p-A scattering, the still outstanding question is what is the mechanism of thermalization of the hadronic system produced in the collision~\cite{Baier:2000sb,Berges:2020fwq,Du:2020dvp,Schenke:2021mxx,Mueller:2021gxd,Fu:2021jhl,BarreraCabodevila:2022jhi,Muller:2022htn}. On the one hand in heavy ion collision hydrodynamic simulations point to very short thermalization times, and on the other hand in p-A collisions, the system appears to show collective behavior even though the number of particles in the final state is not that large~\cite{CMS:2015yux,PHENIX:2016cfs,ATLAS:2019pvn,ALICE:2019zfl,STAR:2019zaf,PHENIX:2018lia}. One interesting idea that has been put forward is the so-called eigenstate thermalization~(see \cite{Deutsch_2018} for review), i.e. manifestation of quasi-thermal properties due to the structure of highly excited states produced in the collision rather than the physical process of thermalization via collisions and/or radiation. Although no quantitative theory based on this hypothesis has been advanced so far, the idea itself is extremely interesting and may yet prove to be fruitful in our understanding of these complex systems. In this context, it is clear that entanglement properties of various degrees of freedom in such excited states are the main driver behind the thermal-like behavior of the interesting observables.

Another possible intriguing connection between quantum entanglement and high energy scattering was proposed in \cite{Kharzeev:2017qzs}. The idea here is that in DIS, one directly probes the content of the wave function of the target (proton) only within a small transverse area $S\sim Q^{-2}$. Hadronic degrees of freedom within this area are entangled within the proton wave function with the rest of the degrees of freedom, particularly with the low momentum slow modes governed by the physics of confinement. This quantum entanglement is ruptured in the scattering event, and the entanglement entropy reflects itself as the Boltzmann entropy of produced particles. Some model calculations to estimate this entropy have been performed~\cite{Zhang:2021hra,Tu:2019ouv,Hentschinski:2022rsa}, and the results semi-quantitatively compared with the data with reasonable agreement. Although the validity of this approach has been questioned in our earlier papers~\cite{Duan:2020jkz,Duan:2021clk}, the possible importance of quantum entanglement in this context remains a very intriguing possibility.

One interesting point related to the previous considerations is the following. In heavy ion collisions, the bulk of the analysis in the hydrodynamic phase is performed in terms of classical hydrodynamics, even though the nature of the effects in question may be due to quantum entanglement. We find ourselves in a somewhat similar situation in the analysis of DIS.  The most interesting regime where quantum entanglement is dominant in the approach of \cite{Kharzeev:2017qzs} is the regime close to saturation, which according to the interpretation of \cite{Kharzeev:2017qzs} leads to a maximally entangled quantum state. On the other hand, this is precisely the regime where the classical description based on the effective theory of Color Glass Condensate (CGC) is supposed to be valid. 

We are thus lead to ask the following question: does classical physics contain some (or possibly many) of the effects of quantum entanglement? 
This is not such an outrageous proposition as it may sound. Consider, for example, one typical measure of entanglement - the entanglement entropy. Quantum entanglement entropy arises in a pure state of an  interacting system, when we integrate out some of the degrees of freedom, thus losing the full information about the system. This loss of information is reflected in the von Neumann entropy of the reduced density matrix. Clearly, one can pose the same question in a classical realm. For an interacting system, we may ask what is the probability distribution of some subset of degrees of freedom if we are not interested in measuring the rest of the DoF's. If all degrees of freedom interact (albeit classically), the distribution of the observed ``system'' depends on the dynamics of the unobserved ``environment''. This classical ``entanglement'' between degrees of freedom  generates entropy on a purely classical level which can be calculated. The question is whether this entropy is related to the von Neumann entropy in the corresponding quantum system.

 In the present paper, we analyze this question in a simple example of an interacting classical (and quantum) system - two coupled harmonic oscillators. The paper is structured as follows. In Sec. II we introduce the notion of classical entanglement for the system of two classical interacting oscillators. We calculate the classical entanglement entropy~\footnote{Note that the commonly used concept of classical entanglement in optics~\cite{Aiello_2015} has no connection to the problem we discuss in this paper.} in the limit where the two oscillators have very different frequencies and the coupling between them is weak. These limits are not  fundamental to the problem, but they allow us to perform the calculation from beginning to end analytically. This also echoes the situation in high energy scattering where naturally one probes the fast degrees of freedom while "integrating over" the slow ones while the interaction between the two is week, since it is governed by the coupling constant on a high momentum scale ($Q$ or  $Q_s$). Our result for the entropy is quite simple and intuitive. In Sec. III, we consider the same two oscillators, but this time in quantum mechanics. We calculate the quantum entropy
of entanglement between the two oscillators for a highly excited eigenstate of the system at weak coupling. In this regime, the calculation can be performed in the WKB approximation completely analytically. We show that the quantum von Neumann entropy of entanglement is the same as the classical entanglement entropy calculated in Sec.II. Finally is Sec. IV, we contrast this result with the entanglement in the ground state of the two oscillators and in an excited state with the energy not high enough to be considered classical. Here we draw a suggestive parallel with the picture of \cite{Kharzeev:2017qzs}, and conclude with a short discussion. Our main conclusion is that the entanglement in the most interesting regime for high energy scattering is most likely of a classical origin and has little to do with genuine quantum effects of the EPR type.

\section{Classical entanglement entropy of two coupled oscillators}
Consider a  system of two classical coupled harmonic oscillators
\begin{align}
\mathcal{H}_c=\frac{1}{2}\left[  p_x^2+p_y^2+\omega^2x^2+\Omega^2 y^2+ 2Cxy  \right]
\end{align}
We consider the situation where the frequencies of the two oscillators are very different, and the coupling is weak:
\begin{equation}
C \ll \omega^2 \ll \Omega^2\,.
\end{equation}
This limit will allow us to study the system analytically.

For some specified initial condition, the system, of course, follows its  trajectory in the phase space, which is the solution of the classical equations of motion
\begin{align}
	\dot{x}&= \frac{\partial \mathcal{H}_c } {\partial p_x} = p_x, \\
	\dot{y}&= \frac{\partial \mathcal{H}_c } {\partial p_y} = p_y, \\
	\dot{p_x}&= - \frac{\partial \mathcal{H}_c } {\partial x}  = - \omega^2 x-Cy, \\
	\dot{p_y}&=  - \frac{\partial \mathcal{H}_c } {\partial y} = - \Omega^2 y - Cx.
\end{align}

Suppose now that we perform measurements only on the coordinate and momentum of the first oscillator $x,\ p_x$. The set of the results of these measurements defines the probability distribution ${\cal W}(x,p_x)$. Since the two degrees of freedom $x$ and $y$ interact with each other, they are classically entangled in the sense that the distribution
${\cal W}(x,p_x)$ depends on the dynamics and the initial conditions of $y,\ p_y$. The extent of this classical entanglement can be quantified by calculating the Boltzmann entropy of ${\cal W}(x,p_x)$ defined as usual as 
\begin{equation}
S_B=-\int dpdx \mathcal{W}(p,x) \ln [\mathcal{W}(p,x)\Delta]\,.
\end{equation}
To define the entropy  we have introduced the minimal phase space volume $\Delta$.
 This is necessary since the classical probability density $\mathcal{W}$ is a dimensional quantity, while in defining entropy one needs to take the logarithm of the probability, which is dimensionless. 
  The dimension has to be canceled by multiplying it by the elementary phase space volume. The existence of such a minimal volume is guaranteed by quantum mechanics, which takes over at small values of the action or equivalently short distances in the phase space. The usual  quantum-classical correspondence requires that we take $\Delta$ to be of the order of the Planck constant. The Heisenberg uncertainty principle would suggest that we use $\Delta\approx\hbar/2$. However, since the classical approximation is not valid for small phase space volumes, we do not believe that the exact value of $\Delta$ can be unambiguously determined "a priori" and then used reliably for comparison of the classical result with the quantum expression that we will derive in the next section. 
 To anticipate the result of the next section, we  will see that the classical and quantum expressions for entropy are identical with the choice $\Delta=h/2$. We will therefore choose this as our definition of the minimal volume $\Delta$\footnote {As we will see in Section III, a different choice  $\Delta=X h/2$  with $X$ a pure number, still leads to quantum and classical entropies equal up to a subleading term. In this sense the exact value of $X$ is not important.} .
  
 Thus even though we are dealing with a classical system, we are forced to  introduce
 the Planck's constant $h$ into the classical expression for entropy.

\subsection{The probability distribution}
We first calculate the probability distribution ${\cal W}(x,p_x)$. 
The two-oscillator system can be easily diagonalized. Define
\begin{align}
x_1=&  \alpha x-     \beta y,  \\
x_2=&\beta x+ \alpha y                                       
\end{align}
and 
\begin{align}
p_1=&\alpha p_x-\beta p_y,          \\
p_2=&\beta p_x +\alpha p_y
\end{align}
with
\begin{equation}
\begin{split}
\alpha^2&=\frac{1}{2}\left[1+\frac{\Omega^2-\omega^2}{\sqrt{(\Omega^2-\omega^2)^2+4C^2}} \right]\approx 1-\frac{C^2}{\Omega^4},\\
\beta^2&= \frac{1}{2}\left[1-\frac{\Omega^2-\omega^2}{\sqrt{(\Omega^2-\omega^2)^2+4C^2}} \right]\approx \frac{C^2}{\Omega^4}\,.
\end{split}
\end{equation}

In terms of these coordinates, the hamiltonian is diagonal. The corresponding frequencies are 
\begin{align}
\omega^2_1=\frac{\omega^2+\Omega^2-\Delta}{2}\quad {\rm and}\quad \omega^2_2=\frac{\omega^2+\Omega^2+\Delta}{2}
\end{align} 
where 
\begin{equation}\Delta=\sqrt{(\Omega^2-\omega^2)^2+4C^2}\approx \Omega^2-\omega^2+\frac{2C^2}{\Omega^2}\,.
\end{equation}
At small $C$, the shift infrequencies is negligible so that $\omega_1^2\approx \omega^2$ and $\omega_2^2\approx \Omega^2$.

The model possesses two conserved quantities 
\begin{align}
E_+=&\frac{1}{2}\left[  p_x^2+p_y^2+\omega^2x^2+\Omega^2 y^2+ 2Cxy  \right],\\
E_-=&2\alpha\beta p_xp_y+ \frac{1}{2}\Big[  (\beta^2\omega_2^2-\alpha^2\omega_1^2) x^2+       (\alpha^2\omega_2^2-\beta^2\omega_1^2) y^2                    \Big] +\alpha\beta(\omega_2^2+\omega_1^2)xy       \\  
 &+\frac{1}{2}     [(\alpha^2-\beta^2)(p_y^2-p_x^2)]   \\ \approx&     \frac{1}{2}\left[p_y^2-p_x^2\right]+\frac{2C}{\Omega^2}p_xp_y+\frac{1}{2}\left[\Omega^2y^2-\omega^2x^2\right]+Cxy   \,.                                        \nonumber
\end{align}
The two are the linear combinations of the energies of two independent harmonic oscillators (in obvious notations)
\begin{align}
E_+&=E_1+E_2,\\
E_-&=E_2-E_1.
\end{align}



The joint classical probability density for the two oscillators is  given simply by the product of the delta functions for each independent oscillator, or for the conserved quantities $E_+$ and $E_-$.
\begin{equation}
\begin{split}
\mathcal{W}(x,p_x; y,p_y)=&N \delta\left(E_+-\frac{1}{2}\left[  p_x^2+p_y^2+\omega^2x^2+\Omega^2 y^2+ 2Cxy  \right]\right)    \\
&\quad \times\delta\left(E_--
\left[\frac{1}{2}\left[p_y^2-p_x^2\right]+\frac{2C}{\Omega^2}p_xp_y+\frac{1}{2}\left[\Omega^2y^2-\omega^2x^2\right]+Cxy\right]
\right)
\end{split}
\end{equation}
where $N$ is the normalization constant determined by the condition
\begin{equation}
\int dydxdp_ydp_x\mathcal{W}(x,p_x; y,p_y)=1\,.
\end{equation}
The ``reduced'' probability density, i.e. the probability density for $x,p_x$ at any $y, p_y$ is given by
\begin{equation}
\begin{split}
\mathcal{W}(x,p_x)=&N \int_{y,p_y}\delta\left(E_+-\frac{1}{2}\left[  p_x^2+p_y^2+\omega^2x^2+\Omega^2 y^2+ 2Cxy  \right]\right)    \\
&\quad \times\delta\left(E_--
\left[\frac{1}{2}\left[p_y^2-p_x^2\right]+\frac{2C}{\Omega^2}p_xp_y+\frac{1}{2}\left[\Omega^2y^2-\omega^2x^2\right]+Cxy\right]
\right)\,.
\end{split}
\end{equation}

For calculational simplicity, we will assume $E_+ \sim E_-$ or $E_1\ll E_2$.  This implies $\frac{p_y}{p_x} \gg 1$. Under this assumption, it is easy to integrate over $y$. We solve the first delta function for $y$ to obtain
\begin{align}
y_{\pm}=-\frac{Cx}{\Omega^2}\pm\frac{1}{\Omega^2}\sqrt{(Cx)^2+\Omega^2[2E_+-p_y^2-p_x^2-\omega^2x^2]}\,.
\end{align}
Integrating over $y$ we then get
\begin{equation}
\begin{split}
\mathcal{W}(x,p_x)=&N \int_{p_y} \delta\left(E_--E_+-\frac{2C}{\Omega^2}p_xp_y+p_x^2+\omega^2x^2\right)
\sum_{\pm} \frac{1}{  |\Omega^2 y_{\pm}+Cx|}\\
=&2 N \int_{p_y} 
\frac{ \delta\left(E_--E_+-\frac{2C}{\Omega^2}p_xp_y+p_x^2+\omega^2x^2\right)}{\sqrt{C^2x^2+\Omega^2(2E_+-(p_x^2+p_y^2+\omega^2x^2))}}\,.
\end{split}
\end{equation}
Since the remaining $\delta$ function is a linear function of $p_y$, it can be easily integrated over with the result
\begin{equation}
\mathcal{W}(x,p_x)=N\frac{\Omega^2}{2C|p_x|} \frac{2}{\sqrt{C^2x^2+\Omega^2(2E_+-(p_x^2+\bar p_y^2+\omega^2x^2))}}
\end{equation}
where 
\begin{align}
\bar p_y=\frac{\Omega^2}{2Cp_x}\left(E_--E_++p_x^2+\omega^2x^2\right)\,.
\end{align}
This is our expression for the probability density of $x,p_x$.
\subsection{The classical entanglement entropy}
The next step is to compute the entropy. 
To facilitate this, we introduce polar coordinates
\begin{equation}
p_x=R\cos\theta; \ \ \ \ \ \ \omega x=R\sin\theta\,.
\end{equation}
We then have
\begin{equation}
\mathcal{W}=2N\left[aR^4+bR^2+c\right]^{-1/2}
\end{equation}
with
\begin{eqnarray}
a&=&-\Omega^2+\frac{4C^2}{\Omega^2}\left(\frac{C^2}{\Omega^2\omega^2}\sin^2\theta-1\right)\cos^2\theta\approx-\Omega^2 -\frac{4C^2}{\Omega^2}\cos^2\theta,\\
b&=&\frac{8C^2E_+}{\Omega^2}\cos^2\theta+2\Omega^2(E_+-E_-)\nonumber,\\
c&=&-\Omega^2(E_+-E_-)^2\,.\nonumber
\end{eqnarray}

The probability distribution can be written in the following simple form
\begin{equation}\label{wx}
\mathcal{W}(x,p_x)\approx \frac{2N}{\Omega}\left[(R^2-X_1)(X_2-R^2)\right]^{-\frac{1}{2}}
\end{equation}
with
\begin{equation}
X_{1,2}=E_+-E_-\mp\frac{2C}{\Omega^2}|\cos\theta|\sqrt{E_+^2-E_-^2}
\end{equation}
where we have only kept terms of order $C$.

Note that the probability density does not vanish only for $X_2\ge R^2\ge X_1$ even though we have not indicated this explicitly in \eqref{wx}. In all the following discussions, the integration over $R^2$ is performed within  these limits\footnote{ Note that
this is consistent with the limit $C\rightarrow 0$, as in this limit $X_1=X_2$ and thus the probability is nonzero only for $p_x^2+\omega^2x^2=E_+-E_-$, which is precisely the energy of the second independent oscillator.}.

First, we calculate the normalization factor $N$ from the condition 
\begin{equation}
\int dp_xdx\mathcal{W}(p_x,x)=1\,.
\end{equation}
Straightforward algebra  leads to 
\begin{equation}
N=\frac{\omega\Omega}{2\pi^2}\,.
\end{equation}

For the entropy, we need to calculate the integral
\begin{equation}
S_{CE}=\int dp_xdx\frac{\omega}{\pi^2}\left[(R^2-X_1)(X_2-R^2)\right]^{-\frac{1}{2}}\ln \left[\frac{ 2\pi^2}{h\omega }\left[(R^2-X_1)(X_2-R^2)\right]^{\frac{1}{2}}\right]\,.
\end{equation}
Using the same polar coordinates, shifting the radial coordinate $R^2=X+E_+-E_-$, and defining $A=\frac{2C}{\Omega^2}|\cos\theta|\sqrt{E_+^2-E_-^2}$ we can simplify this expression to 
\begin{equation}
S_{CE}=\frac{1}{2\pi^2}\int d\theta dX\left[A^2-X^2\right]^{-\frac{1}{2}}\ln\left[\frac{2\pi^2}{h\omega}\left[A^2-X^2\right]^{\frac{1}{2}}\right]\,.
\end{equation}
Finally rescaling $y=X/A$ we obtain
\begin{eqnarray}
S_{CE}&=&\frac{1}{2\pi^2}\int d\theta dy\left[1-y^2\right]^{-\frac{1}{2}}\ln\left[\frac{|A|2\pi^2}{h\omega}\left[1-y^2\right]^{\frac{1}{2}}\right]\\
&=&\frac{1}{2\pi}\int_{-1}^{1} dy \left[1-y^2\right]^{-\frac{1}{2}}\ln\left[1-y^2\right]+\frac{1}{2\pi}\int_0^{2\pi} d\theta \ln|\cos\theta|+\ln\left[\frac{4\pi^2C}{\Omega^2}\frac{\sqrt{E_+^2-E_-^2}}{h\omega}\right]\nonumber\\
&=&\frac{1}{\pi}\int_0^{2\pi} d\theta \ln|\cos\theta|+\ln\left[\frac{4\pi^2C}{\Omega^2}\frac{\sqrt{E_+^2-E_-^2}}{h\omega}\right]\nonumber\,
\end{eqnarray}
where in the last line, we have performed the change of variables $y=\sin\theta$.
Using (Gradshtein-Ryzhik, eq. BI306)
\begin{equation}
\int_0^{\pi/2}\ln\cos x dx=-\frac{\pi}{2}\ln 2
\end{equation}
we obtain finally
\begin{equation}\label{ce}
S_{CE}=\ln\left[\frac{\pi^2C}{\Omega^2}\frac{\sqrt{E_+^2-E_-^2}}{h\omega}\right]=\ln\left[\frac{\pi C}{\Omega}\frac{\sqrt{E_1E_2}}{\hbar\omega\Omega}\right]\,.
\end{equation}

This is our final result for the classical entanglement entropy. Note that 
 formally the entropy is negative for sufficiently small $C$. This indicates that for very small $C$ our classical calculation is not a good approximation to the quantum entanglement entropy. The situation here is similar to the thermodynamic entropy, which is formally negative at low temperatures. We, therefore, expect  that the appropriate regime in which the present classical calculation can be a reliable representation of the true quantum result is 
 when the ratio $E/\hbar\omega$  is large enough (at fixed $C$) so that the entropy is positive. The condition for the positivity of entropy is simply the requirement that the interaction energy of the two oscillators (which is proportional to $C$) is much larger than $\hbar \Omega$, which is very natural from the point of view of the classical limit of quantum mechanics.

 We now turn to calculating the quantum entanglement entropy in the quantized system.


\section{Quantum entanglement entropy of coupled oscillators in a highly excited state} 

Our goal in this section is to calculate the quantum entanglement entropy for the same system (albeit this time quantized) in a highly excited state. 
 More specifically, we take a state in which the energies of both oscillators are high. In addition, we also assume that the interaction energy, although in general suppressed by the factor of $C$, is still large on the quantum scale, i.e.  we assume
\begin{equation}
E_1\gg\hbar\Omega,\ \ \ \ E_2\gg\hbar\Omega; \ \ C^2 \langle x^2y^2 \rangle \sim C^2\frac{E_1E_2}{\Omega^2\omega^2}\gg \hbar^2 \Omega^2\,.
\end{equation}
In this regime, we can use the WKB approximation for excited states wave functions.

We also assume that the interaction energy is smaller than the energy of the decoupled oscillators, i.e.
\begin{equation}\label{csmall}
C\frac{\sqrt{E_1E_2}}{\Omega\omega}\ll \rm{min}\{E_1,E_2\}\,.
\end{equation}

For a single harmonic oscillator, the highly excited state wave function is given by (we take $m=1$ consistently with our definition of the Hamiltonian) is
\begin{align}\label{wkb}
\phi_{n,\ {\rm odd}}^{\rm WKB}(x)=\sqrt{\frac{4}{T p(x)}} \sin( \frac{S(x)}{\hbar}),\ \ \ \phi_{n,\ {\rm even}}^{\rm WKB}(x)=\sqrt{\frac{4}{T p(x)}} \cos( \frac{S(x)}{\hbar})\,.
\end{align}
where $T=\frac{2\pi}{\omega}$ is the period of motion and $S(x)$ is the classical action corresponding to energy $E_n=n\hbar \omega$\footnote{We have ignored the phase inside the $\sin$ and $\cos$ functions, as well as the zero point energy in the expression for the energy eigenvalue, as those are irrelevant for large $n$.}.
\begin{align}
S(x)=\int_0^x dz  p(z)=\sqrt{2} \int_0^x dz \sqrt{E_n-\frac{1}{2}\omega^2 z^2}\,.
\end{align}
with ``classical'' momentum $p$ defined by
\begin{equation}
p(x)\equiv\sqrt{2\left(E-\frac{1}{2}\omega^2 x^2\right)}\,.
\end{equation}
The action can be written as 
\begin{equation}
S(x)=n(\theta+\frac{1}{2}\sin(2\theta)) ; \ \ \ \  \sin(\theta)=\sqrt{\frac{\hbar\omega }{2n}}x\,.
\end{equation}
When $x$ is far away from the classical turning points $|x|_{\rm turn}=\sqrt{\frac{2n+1}{\hbar\omega}}$, we have $\theta\ll 1$ and we get 
\begin{equation}\label{expanded}
S(x)\approx x\sqrt{2n\hbar \omega}\,.
\end{equation}

The $odd$ and $even$ subscripts indicate that the appropriate expressions are valid for odd and even values of $n$.



For the coupled oscillators of the previous section, the eigenfunctions are, of course, just the products of the WKB functions for the two independent oscillators $x_1$ and $x_2$. For definiteness, we choose both occupation numbers to be negative:
\begin{equation}\label{wkb2}
\phi_{n,m}^{\rm WKB}(x,y)=\sqrt{\frac{16}{T_1 T_2p(x_1) p(x_2)}} \sin \left(\frac{1}{\hbar}S_1(x_1)\right)\sin \left(\frac{1}{\hbar}S_2(x_2) \right)
\end{equation}
where $S_1$( and $S_2$) is the action of a single harmonic oscillator with the frequency $\omega$ (and $\Omega$).

To make further progress, we use the fact that $C/\Omega^2\ll 1$ and $\Omega\gg \omega$. We concentrate on the phase of the wave function \eqref{wkb2}. We first expand the action to the first order in the small parameter $\beta$
\begin{equation}\label{s1}
\begin{split}
S_1(\alpha x -\beta y) \approx S_1(x )-\beta y p(x )\, ,
\end{split}
\end{equation}
\begin{equation}\label{s2}
\begin{split}
S_2(\alpha y +\beta x) \approx S_2(y )+\beta x p(y )\,. 
\end{split}
\end{equation}

We now note that although both \eqref{s1} and \eqref{s2} have corrections of order $\beta$, we do not need to keep them both when the frequencies are very different. We can estimate the order of magnitude of the two corrections, by remembering that to leading order in $\beta$,
\begin{equation}
\langle x^2\rangle\sim E_1/\omega^2; \ \ \ \langle y^2\rangle\sim E_2/\Omega^2; \ \ \ \langle p_x^2\rangle\sim E_1; 
 \ \ \ \langle p_y^2\rangle\sim E_2\,.
\end{equation}
We therefore have
\begin{equation}
xp_y\sim\frac{\sqrt{E_1E_2}}{\omega}; \ \ \ yp_x\sim\frac{\sqrt{E_1E_2}}{\Omega};\ \ \ \ \ xp_y\gg yp_x\,.
\end{equation}
Neglecting the smaller of the two, we arrive at the approximation of the wave function as
\begin{eqnarray}\label{wkb3}
\phi_{n,m}^{\rm WKB}(x,y)&\approx&\sqrt{\frac{16}{T_1 T_2p_x p_y}} \sin \left(\frac{1}{\hbar}S_1(x)\right)\sin \left(\frac{1}{\hbar}(S_2(y) +\beta xp_y)\right)\\
&=&\sqrt{\frac{16}{T_1 T_2p_x p_y}} \sin \left(\frac{1}{\hbar}S_1(x)\right)\left[\sin \frac{1}{\hbar}S_2(y)\cos \frac{1}{\hbar}\beta xp_y+\cos \frac{1}{\hbar}S_2(y)\sin \frac{1}{\hbar}\beta xp_y\right]\nonumber\
\end{eqnarray}
with
\begin{equation}
p_x=\sqrt{2[E_1-\frac{\omega^2}{2} x^2]}; \ \ \ p_y=\sqrt{2[E_2-\frac{\Omega^2}{2} y^2]}; \ \ \ \ \ \ E_1=\hbar n\omega; \ \ \ \ E_2=\hbar m\Omega\,.
\end{equation}
Note that in the regime interesting for us we have $\frac{1}{\hbar}\beta xp_y\sim\frac{1}{\hbar} \frac{C\sqrt{E_1E_2}}{\Omega^2\omega}>1$, and therefore we are not allowed to expand the $\sin$ function in \eqref{wkb3} in  $\beta$. On the other hand, the factors $p(x_1)$ and $p(x_2)$ in the prefactor in \eqref{wkb2} are indeed expandable in $\beta$ and we have kept only the leading order term in this expansion in \eqref{wkb3}.

Our goal now is to calculate the reduced density matrix for the $x$ coordinate by integrating out $y$.
We first rewrite the wave function in the following simple form
 \begin{eqnarray}
 \label{wkb4}
\phi_{n,m}^{\rm WKB}(x,y)&=&\sqrt{\frac{4}{T_1 T_2p_x p_y}}\left\{\left[ \sin \left(\frac{1}{\hbar}[S_1(x)+\beta xp_y]\right){+}\sin \left(\frac{1}{\hbar}[S_1(x)-\beta xp_y]\right)\right]\right.
\sin \frac{1}{\hbar}S_2(y)\nonumber\\
&&\left.
-\left[ \cos \left(\frac{1}{\hbar}[S_1(x)+\beta xp_y]\right)-\cos \left(\frac{1}{\hbar}[S_1(x)-\beta xp_y]\right)\right]
\cos\frac{1}{\hbar}S_2(y)\right\}\,.
\end{eqnarray}
The matrix element of the reduced density matrix in coordinate basis is given by
\begin{equation}
\hat\rho(x,\bar x)=\int dy \phi_{n,m}^{\rm WKB*}(\bar x,y)\phi_{n,m}^{\rm WKB}( x,y)\,.
\end{equation}
The integral over $y$ here extends between the classical turning points of the oscillator with frequency $\Omega$ and energy $E_2$.

We now observe that when integrating over $y$ the cross terms involving the product $\sin\frac{1}{\hbar}S_2(y)\cos\frac{1}{\hbar}S_2(y)$ vanish due to parity. In addition we write $\sin^2\frac{1}{\hbar}S_2(y)=\frac{1}{2}\left[1-\cos\frac{2}{\hbar}S_2(y)\right]$ and $\cos^2\frac{1}{\hbar}S_2(y)=\frac{1}{2}\left[1+\cos\frac{2}{\hbar}S_2(y)\right]$ . Since the period of oscillation in $y$ is very short, the terms involving  $\cos\frac{2}{\hbar}S_2(y)$ are suppressed by a factor $1/m$, and for large $m$ can be neglected. 
We, therefore, arrive to the following simple form of the density matrix
\begin{eqnarray}
&&\hat\rho(x,\bar x)^{\rm WKB}\approx\frac{2}{T_1 T_2 p_y}\frac{1}{\sqrt{p_xp_{\bar x}}}\\ \notag 
&&\Big\{\left[ \sin \left(\frac{1}{\hbar}[S_1(x)+\beta xp_y]\right)+\sin \left(\frac{1}{\hbar}[S_1(x)-\beta xp_y]\right)\right] \\&&  \notag   \times \left[ \sin \left(\frac{1}{\hbar}[S_1(\bar x)+\beta \bar xp_y]\right)+\sin \left(\frac{1}{\hbar}[S_1(\bar x)-\beta \bar xp_y]\right)\right]
\nonumber\\\notag 
&&+\left[ \cos \left(\frac{1}{\hbar}[S_1(x)+\beta xp_y]\right)-\cos \left(\frac{1}{\hbar}[S_1(x)-\beta xp_y]\right)\right] 
\\&&  \notag   \times 
\left[ \cos \left(\frac{1}{\hbar}[S_1(\bar x)+\beta \bar xp_y]\right)-\cos \left(\frac{1}{\hbar}[S_1(\bar x)-\beta \bar xp_y]\right)\right]\Big\}\nonumber\,.
\end{eqnarray}
This form is very suggestive. Let us flesh it out. We define 
\begin{equation}
\delta n=\frac{1}{\hbar}\sqrt{\frac{2n}{\omega}}\beta p_y\,.
\end{equation}
Remembering our estimate for $p_y\sim \sqrt {2E_2}$, we see that parametrically
\begin{equation}\delta n\sim \frac{C}{\Omega}\frac{\sqrt{E_1E_2}}{\Omega\hbar\omega}\,.
\end{equation}

Given our assumption \eqref{csmall} we have $\delta n\ll n$. 
We will now use the expanded form of the classical action~\eqref{expanded}. 
We can then write 
\begin{eqnarray}
&&\hat\rho^{\rm WKB}(x,\bar x)=\int dy\frac{2}{T_1 T_2 p_y}\frac{1}{\sqrt{p_xp_{\bar x}}}\\
&&
\left\{\left[\sin\frac{1}{\hbar}x\sqrt{2(n+\delta n)\omega}+\sin\frac{1}{\hbar}x\sqrt{2(n-\delta n)\omega}\right]\left[\sin\frac{1}{\hbar}\bar x\sqrt{2(n+\delta n)\omega}+\sin\frac{1}{\hbar}\bar x\sqrt{2(n-\delta n)\omega}\right]\right.\nonumber\\
&&\left.+\left[\cos\frac{1}{\hbar}x\sqrt{2(n+\delta n)\omega}-\cos\frac{1}{\hbar}x\sqrt{2(n-\delta n)\omega}\right]\left[\cos\frac{1}{\hbar}\bar x\sqrt{2(n+\delta n)\omega}-\cos\frac{1}{\hbar}\bar x\sqrt{2(n-\delta n)\omega}\right]\right\}\nonumber\,.
\end{eqnarray}

It is convenient to change the integration variable from $y$ to $\delta n$:
\begin{equation}
\frac{1}{p_y}dy=\frac{C\sqrt{E_1}}{\sqrt{2}\hbar\omega\Omega^3}\frac{1}{\sqrt{\frac{4E_1E_2C^2}{\hbar^2\omega^2\Omega^4}-(\delta n)^2}}d\delta n\,.
\end{equation}
 Let us define orthonormal  functions
 \begin{eqnarray}
 \phi^1_{\delta n}&\equiv&\sqrt{\frac{1}{T_1p(x) }} \left[\sin \left(\frac{1}{\hbar}S_{n+\delta n}(x)\right)+\sin \left(\frac{1}{\hbar}S_{n-\delta n}(x)\right)\right]\\
 &\approx&\sqrt{\frac{1}{T_1p(x) }} \left[
 \sin\frac{1}{\hbar}x\sqrt{2(n+\delta n)\omega}+\sin\frac{1}{\hbar}x\sqrt{2(n-\delta n)\omega}
\right]\nonumber\\
 \phi^2_{\delta n}&\equiv&\sqrt{\frac{1}{T_1p(x) }} \left[\cos \left(\frac{1}{\hbar}S_{n+\delta n}(x)\right)-\cos\left(\frac{1}{\hbar}S_{n-\delta n}(x)\right)\right]\nonumber\\
 &\approx&\sqrt{\frac{1}{T_1p(x) }} \left[
 \cos\frac{1}{\hbar}x\sqrt{2(n+\delta n)\omega}-\cos\frac{1}{\hbar}x\sqrt{2(n-\delta n)\omega}
\right]\nonumber
 \end{eqnarray}
 for every positive $\delta n$

We can then write
\begin{equation}\label{rhol}
\hat\rho^{\rm WKB}(x,\bar x)=\int_0^{ \sqrt{\frac{4E_1E_2C^2}{\omega^2\Omega^4}}}d(\delta n) \lambda(\delta n) \sum_i\phi^{i*}_{\delta n}(\bar x)\phi^i_{\delta n}( x)
\end{equation}
with
\begin{equation}
\lambda (\delta n)=X\frac{1}{\sqrt{\frac{4E_1E_2C^2}{\hbar^2\omega^2\Omega^4}-(\delta n)^2}}\,.
\end{equation}

The constant $X$ has to be determined so that $\lambda$ is the proper probability density, i.e. 
\begin{equation}
2\int_0^{ \sqrt{\frac{4E_1E_2C^2}{\hbar^2\omega^2\Omega^4}}} d(\delta n)\lambda (\delta n)=1\,.
\end{equation}
This leads to  the following expression 
\begin{equation}
\lambda (\delta n)=\sqrt{\frac{\hbar^2\omega^2\Omega^4}{4\pi^2C^2E_1E_2}}\frac{1}{\sqrt{1-\frac{\hbar^2\omega^2\Omega^4}{4E_1E_2C^2}(\delta n)^2}}\,.
\end{equation}

When written in this form, the density matrix is elementary. In fact, $\lambda(\delta n)$ are exactly the eigenvalues of the density matrix. Note that they are all of parametrically in the same order. The entanglement entropy is quite simple to write
\begin{eqnarray}
S_E&=&-2\int_0^{ \sqrt{\frac{4E_1E_2C^2}{\hbar^2\omega^2\Omega^4}}} d(\delta n)\lambda(\delta n)\ln\lambda(\delta n)=\ln \frac{2\pi C\sqrt{E_1E_2}}{\hbar\omega\Omega^2}+
\frac{2}{\pi}
\int_{0}^1 dy\frac{1}{\sqrt{1-y^2}}\ln \sqrt{1-y^2}\nonumber\\
&=&\ln \left[\frac{\pi C}{\Omega}\frac{\sqrt{E_1E_2}}{ \hbar\omega\Omega}\right]\,.
\end{eqnarray}
We observe that this is identical to the classical entropy calculated in the previous section. As noted in Section II although the {\bf exact} equality is contingent on our choice of the unit of the phase space volume in the classical calculation, even with a different choice, the leading logarithmic term calculated classically and quantum mechanically is the same.


\section{Discussion}

In this note, we have discussed a simple example of  ``entangled'' classical state of two coupled harmonic oscillators. We have calculated the entanglement entropy on the classical theory \eqref{ce} for the weak coupling case. 

We have then considered the quantized system and have calculated the von Neumann entanglement entropy for a highly excited state. We calculated this entropy at weak coupling, assuming that the energy of the interaction of the two oscillators is large. We found that in this regime, the quantum and classical entanglement entropies are equal. 

The reduced density matrix in the quantum theory \eqref{rhol} has an interesting property. It describes a mixed ensemble where the probability of finding a state with occupation number between $n-\Delta n$ and   $n+\Delta n$ with $\Delta n= \sqrt{\frac{4E_1E_2C^2}{\hbar^2\omega^2\Omega^4}}$  is practically independent of the occupation number itself, and vanishes for occupation numbers outside this range.  Thus this is very close to a ``maximally entangled'' state. 

We thus observe that a maximally entangled state in this model has a classical origin. This may sound paradoxical at first sight, but upon some reflection, it may not be so surprising. After all, there is nothing intrinsically quantum in a maximally entangled state, and one can achieve maximal entanglement in the classical regime. We believe this property transcends our simple example of coupled harmonic oscillators and is true also for many other systems\footnote{This is not to say that {\it all} maximally entangle states are classical, as exemplified by the Bell states.}.

It is interesting to compare our result for the entropy with that of the ground state of the same system of coupled oscillators. 
The ground state of coupled oscillators is Gaussian which enables one to compute the entanglement entropy exactly (see e.g. Sec. 8.4.1 of  \cite{IJpelaar:2021oyw}).  
The result is
\begin{equation}\label{vac}
S_E^0=-(1-f)\ln(1-f)-f\ln f
\end{equation}
where for small $C$
\begin{align}\label{f}
f = \frac{C^2}{4 \omega  \Omega^3}=\frac{C^2 E_1E_2}{(\hbar\omega)^2  \Omega^4}
\end{align}
with $E_1 = \frac{\hbar\omega}{2}$, and $E_2 = \frac{\hbar\Omega}{2}$. 

The origin of this expression is very simple. After integrating over the coordinate $y$ one finds that the reduced density matrix for the oscillator $x$ contains only two states - the vacuum of the first oscillator with probability $p_0=1-f$ and the first excited state with probability $f$. This leads to the entropy \eqref{vac}.

Interestingly, for a highly excited state in the quantum regime, i.e. when the energies of the two oscillators are large, but the energy of the interaction is ``quantum'' ( i.e. of order $\hbar\Omega$) we find a similar result. This calculation is easy to perform. We start with our expression for the WKB wave function \eqref{wkb3} and expand it to leading order in $C$ (which is appropriate for small interaction energy)
\begin{eqnarray}\label{wkbs}\notag
\phi_{n,m}^{\rm WKB}(x,y)&=&\sqrt{\frac{16}{T_1 T_2p_x p_y}} \sin \left(\frac{1}{\hbar}S_1(x)\right)\left[\sin \frac{1}{\hbar}S_2(y)\cos \frac{1}{\hbar}\beta xp_y+\cos \frac{1}{\hbar}S_2(y)\sin \frac{1}{\hbar}\beta xp_y\right]\\
&\approx&\sqrt{\frac{16}{T_1 T_2p_x p_y}} \sin \left(\frac{1}{\hbar}S_1(x)\right)\notag \\ &\times& \left[\left(1-\frac{1}{2}. \left( \frac{1}{\hbar}\beta xp_y\right)^2\right)\sin \frac{1}{\hbar}S_2(y)+\frac{1}{\hbar}\beta xp_y\cos \frac{1}{\hbar}S_2(y) \right]\,.
\end{eqnarray}
Now we calculate the reduced density matrix to order $C^2$:
\begin{eqnarray}
\hat\rho(x,\bar x)&=&\sqrt{\frac{16}{T^2_1 p_xp_{\bar x} }}\left[1-Ax^2\right]\sin \left(\frac{1}{\hbar}S_1(x)\right)\left[1-A\bar x^2\right]\sin \left(\frac{1}{\hbar}S_1(\bar x)\right)\nonumber\\
&+&\sqrt{\frac{16}{T^2_1 p_xp_{\bar x} }}B\left[x \sin \left(\frac{1}{\hbar}S_1(x)\right)\right]\left[\bar x\sin \left(\frac{1}{\hbar}S_1(\bar x)\right)\right]
\end{eqnarray}
with
\begin{eqnarray}
A&=&\frac{1}{2}\left(\frac{C}{\hbar\Omega^2}\right)^2\frac{2\Omega}{\pi}\int_{-y_{\rm turn}}^{y_{\rm turn}} dyp_y\sin \frac{1}{\hbar}S^2_2(y)=\frac{1}{2} \left(\frac{C}{\hbar\Omega^2}\right)^2\frac{E_2}{\sqrt 2}
; \nonumber\\
B&=&\left(\frac{C}{\hbar\Omega^2}\right)^2\frac{2\Omega}{\pi}\int_{-y_{\rm turn}}^{y_{\rm turn}} dyp_y\cos \frac{1}{\hbar}S^2_2(y)= \left(\frac{C}{\hbar\Omega^2}\right)^2\frac{E_2}{\sqrt 2}\,.
\end{eqnarray}
Here
\begin{equation}
y_{\rm turn}=\frac{\sqrt{2E_2}}{\Omega}\,.
\end{equation}
Just like in the case of the vacuum state, this reduced density matrix is written as the mixture of two orthonormal states
\begin{equation}
\hat\rho(x,\bar x)=(1-F)\phi_e(x)\phi_e(\bar x)+F\phi_o(x)\phi_o(\bar x)
\end{equation}
where 
\begin{equation}
\phi_e(x)=\xi \sqrt{\frac{4}{T_1p_x}}\left[1-Ax^2\right]\sin \left(\frac{1}{\hbar}S_1(x)\right); \ \ \ \ \phi_o(x)=\zeta\sqrt{\frac{4}{T_1p_x}}x\sin \left(\frac{1}{\hbar}S_1(x)\right)\,.
\end{equation}
To find $F$ we only need to calculate the normalization constant $\zeta$, for which we find
\begin{equation}
\zeta^2=\frac{\omega^2}{\sqrt{2}E_1}
\end{equation}
and thus
\begin{equation}
F=\frac{B}{\zeta^2}=\frac{C^2E_1E_2}{\hbar^2\omega^2\Omega^2}=f\,.
\end{equation}
For very small $C$ such that $F\ll 1$, we, therefore, find for entropy the result identical to the vacuum \eqref{vac}.

A similar factor, $\frac{C\sqrt{ E_1E_2}}{ \hbar\omega \Omega^2}=f^{1/2}$\, appears in our expression for classical entanglement entropy. However, while the result for the vacuum  \eqref{vac} is valid for $f\ll 1$, the classical result holds when $f\gg 1$. As we have discussed above, in the classical regime, the reduced density matrix is a mixture of many $(\sim \Delta n$) excited states which all appear with small and (almost) equal probabilities $p_m\sim f^{-1/2}$. The entropy correspondingly is
\begin{equation}
S_E^n=\sum_mp_m\ln p_m\approx \ln f^{1/2}\,.
\end{equation}

Some of our results here are reminiscent of the discussion in \cite{Kharzeev:2017qzs}. In particular, if we identify $f$ with the ``parton PDF'' of the parton model, in the perturbative regime, $f\ll 1$, the result \eqref{vac} is the same as posited in that paper. The classical regime of our model is then naturally identified with the perturbative saturation regime in DIS, where the coupling is small, $C/\Omega^2\ll 1$,  but the number of excitations (partons) is large, $f\gg 1$. Here we find a result similar to the conjecture of \cite{Kharzeev:2017qzs}, but with the additional power $1/2$ under the logarithm. Interestingly, since in our calculation, this result for entropy is purely classical, we expect that the same is true in QCD. That is the regime where the proton wave function is ``maximally entangled'' is the regime of classical gluon fields, and the entanglement itself is purely classical in the sense discussed in the present paper.

 We are tempted to go even further in this analogy and suggest that the situation is similar to the quantum thermalization hypothesis. That is, the quasi-thermal entanglement entropy, as well as other thermal-like properties of highly excited states, are indeed a manifestation of classical entanglement originating from the dynamics of the classical ergodic system. This, of course, cannot be strictly concluded from the example considered in the present paper but seems to us a possibility well worth exploring.

\acknowledgments 
We thank Hassan El Saed for the discussions at the early stages of this work. A.K. and V.S.  thank M.~Lublinsky (Ben-Gurion University of the Negev)  for the support and hospitality in January 2023 when this paper was finalized.  

A.K. is supported by the NSF Nuclear Theory grant 2208387. This material is based
upon work supported by the U.S. Department of Energy, Office of Science, Office of Nuclear
Physics through the Contract No. DE-SC0020081 (H.D. and V.S.) and the Saturated Glue (SURGE) Topical Collaboration (A.K. and V.S.). 


\bibliography{Paper.bib}

\end{document}